\documentclass[12pt]{article}
\usepackage{etex}
\usepackage{epsfig}
\usepackage{epsf}
\usepackage{latexsym}
\usepackage{amsfonts,amssymb,amsmath} 
\usepackage{mathtools}
\usepackage{color}
\usepackage[a4paper,vmargin=3cm,hmargin=2.1cm]{geometry}
\usepackage{tikz}
\usepackage[all]{xy}
\epsfverbosetrue
\newcommand{\hh}{\mathcal{H}}

\newcommand{\de}{\hbox{\rm{d}}}

\newcommand{\pa}{\partial}

\newcommand{\lb}{\left[}
\newcommand{\rb}{\right]}
\newcommand{\lp}{\left(}
\newcommand{\rp}{\right)}
\newcommand{\la}{\left\{}
\newcommand{\ra}{\right\}}
\newcommand{\ul}[1]{\underline{#1}}

\newcommand{\dpp}{\vcentcolon}
\newcommand{\bb}{\begin{eqnarray}}
\newcommand{\ee}{\end{eqnarray}}
\newcommand{\eee}{\nonumber\end{eqnarray}}
\newcommand{\qq}{\quad}

\newcommand{\nc}{\newcommand}
 \nc{\alf}{\alpha} \nc{\La}{\Lambda}
  \nc{\ze}{\zeta}
\nc{\tht}{\theta} \nc{\T}{\Theta} \nc{\be}{\beta}  \nc{\eps}{\epsilon} 
\nc{\ga}{\gamma}  \nc{\De}{\Delta} 
 \nc{\G}{\Gamma}  \nc{\vphi}{\varphi}
 \nc{\si}{\sigma}  \nc{\ka}{\kappa}   \nc{\Si}{\Sigma} 
\nc{\om}{\omega}  
\nc{\chic}{\widehat{\chi}}

\nc{\qqq}{\quad\quad}               

 \nc{\Om}{\Omega}
\nc{\nf}{\infty}   \nc{\dl}{\mathop{\smash{\cal L}}}  \nc{\black}{\rule{3mm}{3mm}}
\nc{\ol}{\overline}        \nc{\und}{\underline} 
\nc{\beq}{\begin{equation}}  \nc{\eeq}{\end{equation}}  \nc{\pt}{\partial}  
   \nc{\dst}{\displaystyle}  \nc{\na}{\nabla} 
\nc{\nnb}{\nonumber}    \nc{\bs}{\backslash}        \nc{\mb}{\mathbb}   
\nc{\sn}{{\rm sn}\,} \nc{\cn}{{\rm cn}\,}     \nc{\dn}{{\rm dn}\,} \nc{\nin}{\noindent}
\nc{\ti}{\tilde}   \nc{\wti}{\widetilde}   \nc{\h}{\hat}  \nc{\wh}{\widehat}
\nc{\tpsi}{\wti{\psi}}   \nc{\tphi}{\wti{\phi}}  \nc{\tH}{\wti{H}} 
\nc{\arc}{{\rm arc}\,}
\newcounter{muni}
\newenvironment{remunerate}{\begin{list}{{\rm \arabic{muni}.}}
{\usecounter{muni}
\setlength{\leftmargin}{0pt}\setlength{\itemindent}{38pt}}}{\end{list}}

\nc{\brm}{\begin{remunerate}}   \nc{\erm}{\end{remunerate}}

\nc{\stg}{\mathop{\smash{*}}}
\nc{\st}{\mathop{\smash{\delta}}}
\nc{\barr}{\begin{array}}   \nc{\earr}{\end{array}}   \nc{\dg}{\dagger}
\nc{\mtvb}{\mathversion{bold}}   \nc{\mtvn}{\mathversion{normal}}

\begin{document}

\thispagestyle{empty}

\begin{center}
${}$
\vspace{3cm}

{\Large\textbf{Axial Bianchi I meets drifting extragalactic sources}} \\

\vspace{2cm}

{\large
Thomas Sch\"ucker\footnote{
Aix Marseille Univ, Universit\'e de Toulon, CNRS, CPT,  Marseille, France\\\indent\qq
thomas.schucker@gmail.com }
}

\vspace{3cm}

{\large\textbf{Abstract}}
\end{center}

We recompute {\it ab initio} the drift of comoving sources in axial Bianchi I universes. Our result resolves the long-standing disagreement between the computations by Quercellini  et al.  from 2009 and by Marcori et al.  from 2018 in favour of the latter. These computations have a
potential impact on Gaia data analysis and cosmological parameter estimation.

\vspace{7cm}

Key-Words: cosmological parameters 

\vspace{0.3cm}
2407.02889v2

\section{Introduction}

As early as 1933, Georges Lema{\^i}tre \cite{lemaitre}, following a recommendation by Albert Einstein, considers (tri-axial) Bianchi I universes without using this name. They generalize flat Robertson-Walker universes by allowing three different scale factors in three orthogonal directions:
\bb 
 \de \tau^2 = \de t^2 - a(t)^2\de x^2-b(t)^2
\de y^2- c(t)^2\,\de z^2,\qq a(t),\,b(t),\,c(t)>0\label{3axBianchi}
\ee
These universes are homogeneous and anisotropic with a 3-dimensional Abelian isometry group. Its Lie algebra is the first in the list of Luigi Bianchi's famous classification from 1898 \cite{Bianchi} of the real, 3-dimensional Lie algebras. In the same work he also 
\begin{itemize}\item 
exhibits 3-dimensional Riemannian manifolds on which these Lie algebras act transitively as Killing vectors, 
\item
he exhibits 3-dimensional Riemannian manifolds on which 4-dimensional Lie algebras act in the same way and shows that they all contain the 3-dimensional Lie algebras from his list as ideals,
\item
and he shows that 5-dimensional Lie algebras cannot act in this way.
\end{itemize}

 For Bianchi I this manifold is simply the 3-dimensional Eulidean space and the manifolds for the other Lie algebras of type Bianchi II - IX underlie the corresponding higher  
Bianchi universes,   whose study began in the late sixties \cite{EMC}-\cite{ch}.

The universes with  four Killing vectors are special cases, which are called locally rotationally symmetric  (L.R.S.) or axial.

The proper time element of the axial Bianchi I universe is obtained by choosing two equal scale factors, e.g. $b=a$. Then there is a fourth Killing vector, $x\pa_y-y\pa_x$, generating rotations around the $z$-axis. 
Note that if all three scale factors are different, the Bianchi I universe has three privileged axes, the $x$-, $y$- and $z$-axis. The axial Bianchi I universe has one privileged axis.
If all three scale factors are equal, we are back at  flat Robertson-Walker universes with the maximally symmetric Euclidean 3-space, three translations, three rotations 
and no privileged axis.

\subsection{Drift of extra-galactic sources in Bianchi I universes}

\begin{itemize}\item
 In 2009 Quercellini et al. \cite{Quer09} compute the drift of comoving  sources as seen by a comoving observer in tri-axial Bianchi I universes. It is given by their equations  (6) and (7). See also Quercellini et al. \cite{Quer10}, equations (33) and (34).
 \item
In 2013 Titov \& Lambert \cite{tl}  measure the drift of 429 extra-galactic radio sources  
 using Very Long Baseline Interferometry.
 \item
 In  2014 Darling \cite{dar} applies the drift equations (6) and (7) by Quercellini et al. to fit the drift of the radio sources observed by Titov \& Lambert. He finds no significant signal.
 \item 
 In 2018 Marcori et al. \cite{Mar} recompute the drift including peculiar velocities of source and observer. Although they do not spell out the drift equations their calculation is in disagreement with  equations (6) and (7) by Quercellini et al..
 \item
Today, thanks to the Gaia mission, we have hundreds of thousands of Quasars with drift and redshift measurements ($z$ up to 4.7) begging to be fitted. 
\end{itemize}

\subsection{Aim and motivation of choices}

    The purpose of this work is resolving the mentioned disagreement and presenting correct drift equations ready to be confronted with the new high-quality Gaia data. Our calculation is  {\it ab initio} and {\it low tech} and concerns the drift of a {\it comoving source} in {\it axial} Bianchi I universes filled with comoving dust and admitting {\it small anisotropy}. We also include an {\it observer's small peculiar velocity}.

\vspace{0.3cm}
Let us comment our choices:
\vspace{0.3cm}
 \begin{description}
\item[ab initio:]
Our only prerequisites are the geodesic equation, Einstein's equation and the definition of cosmic time $t$. 
\item[low tech:]
Our aim is that an experimental colleague can verify all calculations.

As Quercellini et al. we work in coordinates and use  Cartesian coordinates $(x,y,z)$ and their spherical coordinates $(r,\varphi ,\theta )$. 

All approximations are first order Taylor expansions in 
four small 
dimensionless quantities:
\begin{itemize}\item
Light of atomic period $T$ is emitted by an extragalactic source and observed with Doppler-shifted period $T_D$, spectroscopic redshift and incoming direction. We suppose that its time of flight $t_{of}$ is superior to that of light from Andromeda $t_{of} >10^{14}\,$s. Admitting the detection limit of the James Webb Space Telescope, $T_D<10^{13}\,$s, we define the small dimensionless parameter $\eps_s\dpp=T_D/t_{of}<10^{-27}$, $\cdot_s$ for spectroscopic, and use it to compute the redshift by means of a  first order Taylor expansion in $\eps_s$. 
 \item
 The incoming direction is recorded over a tracking time of say 10 years, which we also denote by $T_D$. We define $\eps_d\dpp=T_D/t_{of}<3\cdot10^{-6}$, $\cdot_s$ for drift.
 \item
 We allow the observer to have a small peculiar velocity $V$ divided by the speed of light (that we set to 1).
 We will restrict the fit to $V<10\,\%.$
 \item
 We allow for a small time-dependent anisotropy, $\eta(t)$ with $|\eta(t)|<10\,\%$  between emission and reception today, $t\in[t_e,t_0]$. 
  \end{itemize}
 These four small quantities are considered independent variables, not necessarily of the same order of magnitude.
 The drift equations e.g.\,\,(\ref{dthetlin}) are of the form
 $\delta \theta \sim(a\, \eta+b\,V)\,\eps_d$.
 We denote by the symbol $\sim$  an equality neglecting  terms in $\eps_d^2$, $\eta^2$, $V^2$ or $\eta\,V$ and higher orders, such that each neglected term is less than 1 \%. We suppose that all functions encountered are analytic and acknowledge that the domains of convergence must be controlled during the fit by reducing the domain of input parameters as necessary. In the example we must control that the coefficients $a$ and $b$ remain of order 1.
 
\item[comoving source:]
Peculiar velocities of the sources are not included because they would add three more parameters for each single source. But we may hope that in a large enough sample the peculiar velocities of the sources average out and become less and less important with increasing redshift.
\item[observer's small peculiar velocity:]
A small peculiar velocity of the observer is considered because it only adds three more parameters, which might be relevant for sources with redshifts below 0.5.
 
 \item[Axial] Bianchi universes have four Killing vectors
 while generic Bianchi universes have only three Killing vectors. In this sense the axial Bianchi universes are closer to the  `Robert\-son-Walker' universes that obey the cosmological principle (six Killing vectors) than the generic Bianchi universes. Now the four Killing vectors of axial Bianchi I universes imply geodesics with four conserved quantities, which in turn produce equations for redshift and drift that are much simpler and involve less arbitrary parameters in their fit compared to tri-axial Bianchi I universes. 

  Nevertheless axial Bianchi I universes are sufficient to trace the errors in Quercellini et al.'s calculation.

\item[small anisotropy:] As the cosmological principle with its  six space-like Killing vectors is quite successful in fitting today's data it is a wise choice to remain close to Robertson-Walker universes. For axial Bianchi I universes this means small anisotropy $\eta (t)\dpp={\textstyle\frac{2}{3}} [c(t)/a(t)-1]\rightarrow 0$.
We will be more precise about this parametrization and motivate it at the beginning of section 5.
Note that there are only two other axial universes admitting this limit: axial Bianchi V and IX.
 Axial Bianchi V however is incompatible with Einstein's equation sourced by comoving dust \cite{farn}\cite{val}.
 \end{description}

During our calculation we carefully separate the three parts:
{\it(i) kinematics:}
position of the source on the celestial sphere as a function of the observer's cosmic time and the redshift of the source, {\it(ii) dynamics:} Einstein's equation with cosmological constant and comoving dust, 
{\it(iii) approximation of small anisotropy:} 
$ |\eta(t)|\ll1$.

On the other hand Quercellini et al. mix kinematics and dynamics in their calculation.

\subsection{Summary of results}

Quercellini et al.  use polar angles $\varphi $ and $\theta $ to compute the drift of  a comoving source as seen today $t_0$ by a comoving observer, $V=0$, in tri-axial Bianchi I universes. Neglecting the curvature of the photons' trajectories they find equations (6) and (7) (in \cite{Quer09}) for the $\varphi $- and  $\theta $- components of the drift. Note that these equations are independent of the photon's emission time and of its redshift. It is claimed that the approximation is good to 7 \% for redshifts up to 1, if the Bianchi I universe satisfies Einstein's equations with positive cosmological constant $\Lambda $ and comoving dust today $\rho _0$. 

In the axial case, $b=a$, the $\varphi $-drift vanishes and one is left with equation (7), which becomes independent of $\varphi $.

 For small anisotropy $\eta(t)$  and to linear order in $\eta$, their equation (7) reduces further to
 $$
 \delta \theta\ \sim\   3\sin\theta\cos\theta  \left(-{\textstyle\frac{1}{2}} \, \eta'_0\quad\quad\quad\quad\quad\quad\right)
\,T_D\,,
$$
$T_D$ being the duration of observation.
Our calculation  yields  (equation (\ref{dthetlin}) below with $V=0$),
$$
\delta \theta\ \sim\   3\sin\theta\cos\theta  \left(+{\textstyle\frac{1}{2}} \, \eta'_0+\eta_e/a_{F0}/\chi_{e0}\right)
\,T_D\,,
$$
with the emission time $t_e$ and
$$
\eta_e\dpp=\eta(t_e), \quad \chi_{e0}\dpp=\int_{t_e}^{t_0}1/a_F\dpp=\int_{t_e}^{t_0}1/a_F(t)\,\de t.
$$
The emission time $t_e$ is computed from the {\it direction-dependent} redshift $\underline z_0$ using our equation  (\ref{redshift}) with $V=0$, 
$$
\underline z_0+1\ \sim\ \frac{a_{F0}}{a_{F}(t_e)}
[ 1+ {\textstyle\frac{1}{2}} 
(1-3\cos^2\theta )\eta_e
 ].
 $$
We explain in equation (\ref{4.38}) that for  $\underline z_0=1$ the redshift dependent piece, which is neglected by Quercellini et al., is in absolute value more than four times {\it larger} than the redshift independent piece,  rather than {\it smaller} by a factor of 7 \% as claimed by the authors. Notice that the Gaia catalogue contains objects with redshift up to 4.7.

As shown we also find the opposite sign in front of the redshift independent piece.

Our results and those of Marcori et al. \cite{Mar} agree in their domain of overlap.

\section{The path of a light ray}
 
 The axial Bianchi I metric can be written
 \bb 
 \de \tau^2 = \de t^2 - a(t)^2\lb\de x^2
+\de y^2\rb- c(t)^2\,\de z^2,\qq a(t),\,c(t)>0.\label{bianchi}
\ee
We follow the notations of \cite{stv}. Consider a geodesic $x^\mu (q)$ and denote the derivation with respect to the affine parameter $q$ by $\dot {}\dpp=\de/\de q$. The four Killing vectors $\pa _x,\,\pa _y,\, \pa _z$ and $x\pa _y-y\pa _x$ yield four conserved quantities by Emmy Noether's theorem:
 \bb A\dpp =a(t(q))^2\dot x,\ B\dpp=a^2\dot y,\ C\dpp=c^2\dot z\qq\text{and}\qq Ay-Bx=0.
 \label{noet}
 \ee
The physical significance of the Noether constants $A$, $B$ and $C$ in terms of conserved quantities is summarized in appendix A.
 
 The last equation implies that if $A$ and $B$ both vanish the geodesic is a straight line in the $\pm\, z$-direction. Otherwise the geodesic remains in the plane defined by the $z$-direction and the direction of the initial 3-velocity. 
  
 Let us suppose the geodesic is light-like. Then from $\dot x^\mu g_{\mu \nu }\dot x^\nu=0$ we obtain 
 \bb
 \dot t= \,\frac{1}{W}\, \qq\text{with}\qq 
 W\dpp=
 \lp \frac{A^2+B^2}{a(t)^2}\, + 
\,\frac{C^2}{c(t)^2} \rp ^{-1/2}.
\ee
Eliminating the affine parameter $q$ in favour of the cosmic time $t$ and using $'\dpp=\de/\de t$, the light-like geodesic must satisfy the three first order equations:
\bb 
\vec x'= \,\frac{\dot{\vec x}}{\dot t}\,=\,W
 \begin{pmatrix}
 \,\frac{A}{a^2}\, \\[0.5mm]
   \,\frac{B}{a^2}\, \\[0.5mm]
   \,\frac{C}{c^2}\,
   \end{pmatrix} .\label{vxp}
\ee 
 Let us fix the Noether constants $A,\,B$ and $C$  in terms of the initial velocity of a photon emitted by a comoving source at time $t_e$ and position $\vec x_e\dpp=\vec x(t_e)$. To simplify notations we put $\vec x_{e}=0$.
    
Let us say that the photon arrives today $t=t_0$ in our telescope at position $\vec x_0\dpp=\vec x(t_0)$. Then we have:
\bb
 x_0=\int_{t_{e}}^{t_0}\,\frac{A}{a^2}\,W,\qq y_0=\int_{t_{e}}^{t_0}\,\frac{B}{a^2}\,W,\qq z_0=\int_{t_{e}}^{t_0}\,\frac{C}{c^2}\,W.\label{sol}
 \ee
In our conventions the speed of light is unity; proper time $\tau$, the coordinate time $t$ and the affine parameter $q$ have units of seconds, the comoving coordinates $x,\,y,\,z$ are dimensionless and the scale factors $a$ and $c$ and the Noether constants $A,\,B$ and $C$ carry seconds. Then $W$ is dimensionless. Equation ({\ref{vxp}) remains unchanged under a simultaneous rescaling of the three Noether constants. Therefore we may
choose $N\dpp=\sqrt{A^2+B^2+C^2}$ to be any positive constant.
Through a rescaling of the coordinates $x,\,y,\,z$ we may  achieve $a_0=c_0,$  with $a_0\dpp = a(t_0), \dots$

 We also learn from equations (\ref{vxp}) or (\ref{sol}) that in the Bianchi I metric, the apparent position of comoving sources in the sky changes with time, `cosmic parallax, apparent motion, proper motion or drift' (unless they lie in the $xy$-plane, $C=0$, or on the $z$-axis, $A=B=0$). 
     
\section{A second, infinitesimally close light ray}

Consider now a second photon emitted a short time $T$ later from the same source and arriving in our telescope a time $T_D$ after the first. (The subscript $\cdot_D$ stands for Doppler.) The following calculations are relevant in two physical situations: 
\begin{itemize}\item
$T$ is an atomic period at emission. Then $T_D$ is the Doppler shifted atomic period observed here and today and we define the redshift today
\bb
\underline z_0\dpp=\,\frac{T_D-T}{T}\,.
\ee 
We underline the redshift $\underline z$ to avoid confusion with the coordinate $z$.

If we pretend to know the metric and it has nice properties, then the redshift allows us to compute the time of flight $t_{of}$ of the photon.
\item
$T_D$ is the tracking time during which the observer tracks the source on the celestial sphere.
 Of course, a change in direction can only be measured with respect to another direction that supposedly does not change. Axial Bianchi I universes have indeed the luxury of such fixed directions: the $z$-axis and the $xy$-plane.
\end{itemize}
Let us denote the second geodesic by $\tilde{\vec x}(t)$ with $t$ varying from $t_{\tilde e}\dpp=t_e+T$ to $t_{\tilde 0}\dpp=t_0+T_D$.

In both situations the second photon is sent out
at 
\bb
\tilde{\vec x}(t_{\tilde e})={\vec x}(t_{ e})=\dpp\vec x_{e}=0,
\ee
 because we suppose the source to be comoving. For the observer, on the other hand, we allow for a small peculiar velocity today 
\bb
\vec V=\dpp a_0\,\vec v\,,\qq\qq V<10\,\%.  
\ee
Then the photon arrives in her telescope at position $\vec x_0+T_D\,\vec v$:
\bb
\tilde{\vec x}(t_{\tilde 0})=\dpp\tilde{\vec x}_{\tilde 0}={\vec x}(t_{ 0})+T_D\,\vec v=\vec x_0+T_D\,\vec v.
\ee
  We suppose that  $T_D$ is small with respect to the time of flight 
  $ t_{of}$; so small that we may keep only linear terms in $T$ and $T_D$. Also the peculiar velocity is small enough so that we may keep only linear terms in $V$. In particular, we neglect the dilation of the observer's proper time with respect to cosmic time $t$.
  
  We may use a rotation in the $xy$-plane to bring the peculiar velocity into the form
  \bb
  \vec v=
   \begin{pmatrix}
v_x \\
 0 \\
 v_z
   \end{pmatrix}.
   \ee
Our task is to slightly change the initial direction of the second geodesic $\tilde{\vec x}(t)$ coded in the Noether constants
 $\tilde A,\,\tilde B$ and $\tilde C$, in such a manner that $\tilde{\vec x}_{\tilde 0}=\vec x_0+T_D\,\vec v$. 
 To this end we make the uncommitted Ansatz:
\bb \tilde A
=\dpp
 A+\alpha N,\qq \tilde B
=\dpp
 B+\beta  N,\qq
\tilde C
=\dpp
 C+\gamma N ,\ \ \text{
 implying
 }\ \ 
\alpha ,\,\beta ,\,\gamma \sim 
T_D/t_{of}
\,\dpp=\,\eps.
\ee
Thanks to the invariance under rescaling explained four lines after equation (\ref{sol})
 we may choose $\tilde N=N$ implying 
\bb
A\alpha+B\beta +C\gamma \sim0. 
\label{abc0}
\ee
The second trajectory is computed as the first: 
\bb
\tilde x_{\tilde 0}=\int_{t_{e}+T}^{t_0+T_D}\,\frac{\tilde A}{a^2}\, \tilde W,\qq
\tilde y_{\tilde 0}=\int_{t_{e}+T}^{t_0+T_D}\,\frac{\tilde B}{a^2}\, \tilde W,\qq
\tilde z_{\tilde 0}=\int_{t_{e}+T}^{t_0+T_D}\,\frac{\tilde C}{c^2}\, \tilde W,
\ee
 with 
 \bb \tilde W\!\!\!&\dpp=&\!\!\!
 \lp \frac{\tilde A^2+\tilde B^2}{a^2}\, + 
\,\frac{\tilde C^2}{c^2} \rp ^{-1/2}.
\ee
We linearize $\tilde W$,
 \bb \tilde W\!\!\!&\sim&\!\!\!
W\lb1-W^2N\lp\frac{A\alpha +B\beta }{a^2}\,+\,\frac{C\gamma }{c^2}\rp\rb
\\[2mm]
&\sim& W\lb1+W^2NC\gamma \lp\frac{1}{a^2}\,-\,\frac{1}{c^2}\rp\rb
\sim W\lb 1+\,\frac{N}{C}\,\gamma \lp N^2 \,\frac{W^2}{a^2}\, -1\rp\rb \label{tildeWa}
\\[2mm]
 &\sim& W\!\lb1-W^2N(A\alpha +B\beta ) \!\!\lp\frac{1}{a^2}\,-\,\frac{1}{c^2}\rp\!\rb
\sim \!W\lb 1-\,\frac{NC}{A^2+B^2}\,\gamma\! \lp N^2 \,\frac{W^2}{c^2}\, -1\rp\!\rb, 
\label{tildeW}
\ee
where we have used successively the two identities:
\bb
W^2 \lp\frac{1}{a^2}\,-\,\frac{1}{c^2}\rp=\,\frac{N^2}{C^2}\,\frac{W^2}{a^2}\,-\,\frac{1}{C^2}\,  =-\,\frac{N^2}{A^2+B^2}\,\frac{W^2}{c^2}\,+\,\frac{1}{A^2+B^2}\, .
\ee
(One might think that the details of equations (\ref{tildeWa}) and (\ref{tildeW}) are redundant, but they will be used to compute equations (\ref{z0}) and (\ref{tildex02}) below.)
Then we have to first order in $T/(t_0-t_{e})$:
\bb
\tilde x_{\tilde 0}\!\!\!&\sim&\!\!\!\int_{t_{e}+T}^{t_0+T_D}\lp\frac{ A\tilde W}{a^2}\, +
\,\frac{NW}{a^2}\,\alpha \rp\\[2mm] 
\!\!\!&\sim&\!\!\!
\,\frac{AW_{0}}{a^2_{0}}\, T_D-\,\frac{AW_{e}}{a^2_{e}}\, T
+ \!\!\int_{t_{e}}^{t_0}\la\frac{AW}{a^2}\,-\frac{AW}{a^2}\,\,\frac{NC}{A^2+B^2}\,\gamma\! \lp  \frac{N^2W^2}{c^2}\, -1\rp+\,\frac{NW}{a^2}\,\alpha  \ra
\label{tildex02}
\\[2mm]
\!\!\!&
= &\!\!\!x_0+A \,F\,T_D+\lp\frac{AC}{A^2+B^2}\,\gamma +\alpha \rp I
-\,\frac{AC}{A^2+B^2}\,\gamma J
\ +\ O(\eps^2).
\label{tildex03} 
\ee
In (\ref{tildex02}) we have used the second member of (\ref{tildeW}) and in  (\ref{tildex03}) we have introduced the short hands,
\bb
F\dpp=\,\frac{W_0}{a_0^2}\, -\,\frac{W_e}{a_e^2}\,\frac{T}{T_D}\, ,\qq
F_c\dpp=\,\frac{W_0}{c_0^2}\, -\,\frac{W_e}{c_e^2}\,\frac{T}{T_D}\, ,\qq
I\dpp=N\int_{t_e}^{t_0}\,\frac{W}{a^2}\, ,\qq
J\dpp=N^3\int_{t_e}^{t_0}\,\frac{W^3}{a^2c^2}\,.
\ee 
Since $\tilde x_{\tilde 0}=x_0+v_xT_D$ we have:
\bb 
v_xT_D&\sim& 
A \,F\,T_D+\alpha I
+\,\frac{AC}{A^2+B^2}\,\gamma (I-J). \label{x0}
\ee 
Similarily  for $\tilde y_{\tilde 0}=y_0$ and $\tilde z_{\tilde 0}=x_0+v_zT_D$:
\bb 
0&\sim& 
B \,F\,T_D +\beta I
+\,\frac{BC}{A^2+B^2}\,\gamma (I- J), \label{y0}
\\[2mm]
v_zT_D&\sim& 
C \,F_c\,T_D+ \gamma J. \label{z0}
\ee 
To derive equation (\ref{z0}) we used the second member of (\ref{tildeWa}).
Let us take the following linear combination of the last three equations, $A$(\ref{x0})+$B$(\ref{y0})+$C$(\ref{z0}):
\bb
(Av_x+Cv_z)\,T_D\sim(A^2+B^2)\,F\,T_D + C^2\,F_c\,T_D
=\,\frac{T_D}{W_0}\,-\,\frac{T}{W_e}\,  .
\ee
 From it we obtain the direction-dependent redshift today,
\bb
\underline z_0+1\dpp=\,\frac{T_D}{T}\, 
\,=\,\frac{1}{W_{e}}\,
\frac{1}{1-(A/N\,V_x+C/N\,V_z)}\,
+\ O(\eps_s),
 \label{directz}
 \ee
 since here $T_D$ is an atomic period and $V<1$ may be finite. From now on we ignore $\eps_s$. 
 
We will also need the
 corrections $\alpha ,\,\beta,\,\gamma $ to the second photon's initial direction. To this end let us consider the linear combination $B$(\ref{x0})$-A$(\ref{y0}):
 \bb
 \alpha\,  \sim \,\frac{A}{B}\, \beta+\,\frac{1}{I}\,v_x\,T_D
 . \ee
 Using equations (\ref{abc0}) and (\ref{z0}) we obtain,
 \bb
 \alpha &\sim&\frac{1}{A^2+B^2}\,  \frac{1}{J} \lb AC^2 F_c+B^2\,\frac{J}{I}\,  v_x-AC\,v_z\rb T_D,
  \label{alpha}
 \\[2mm]
 \beta  &\sim&\frac{B}{A^2+B^2}\,  \frac{1}{J} \lb C^2 F_c+A\,\frac{J}{I}\,  v_x-C\,v_z\rb T_D,
 \label{beta }
 \\[2mm]
 \gamma   &\sim& \hspace{17mm}\frac{1}{J} \lb -C\, F_c+v_z\rb T_D.
 \label{gamma}
 \ee

\section{Arriving angles}

Our first task is to compute the angle $\theta $ between the $z$-axis and the velocity $\vec x'_0$ of the first photon. 
This calculation takes place in the tangent space to space-time at $(t_0, \vec x_0)$ with diagonal Lorentz-metric and evident reduction to 3-space. 
 The negative of the spatial part of the metric (\ref{bianchi}) makes this 3-space Euclidean with the standard metric, because $a_0 =c_0$.
  Note that the velocity vector $\vec x'$ is a unit vector with respect to this 3-metric, because we have set the speed of light to one. If we denote by
\bb
\vec u_{z} \dpp=
\begin{pmatrix}
0\\0\\\frac{1}{c} 
\end{pmatrix}
\ee
 the unit vector in the $z$-direction, we have 
\bb
\cos \theta = \vec x'_0\cdot\vec u_{z0}=
W_0\,\frac{C}{c_0^2}\,\, c_0^2\,\,\frac{1}{c_0}\, =\,\frac{CW_0}{c_0}\,=\,\frac{C}{N}\, .
\label{costh}
\ee
\\
 \vspace{4mm}
 \begin{center}
\begin{tabular}{c}
\xy
(0,0)*{}="E";
(0,-5)*{\vec x_e };
(85,35)*{}="O";
(70,25)*{}="Ot";
"E"; "O" **\crv{(40,40)};
"E"; "Ot" **\crv{(44,40)};
{\ar "O"; "Ot"};
{\ar "O"; (85,48)};
(85,40)*{.}; 
(85,53)*{\vec u_{z0}};
{\ar "O"; (96.5,33.5)};
(89.5,34.3)*{.};
(85,40); (89.5,34.3) **\crv{(89,39)};
{\ar "Ot"; (70,40)};
(70,31)*{.};
(70,45)*{\vec u_{z\tilde0} };
{\ar "Ot"; (82,18.5)};
(74.9,22.3)*{.};
(70,31); (74.9,22.3) **\crv{(73,30)};
(30,29)*{\vec x(t) };
(37,20)*{\tilde{\vec x}(t) };
(102,33)*{{\vec x}'_{0} };
(86,15)*{{\tilde{\vec x}}'_{\tilde0} };
(91,39)*{\theta };
(77,25)*{\tilde\theta };
(84.5,30)*{\vec v \,T_D };
(0,-16)*{};
\endxy
\end{tabular}\linebreak\nopagebreak
{Figure 1: Two infinitesimally close geodesics} 
\end{center}
  \vspace{4mm}

Our second task is to compute the same angle $\tilde\theta $ for the second photon at its arrival time $t_{\tilde 0}=t_0+T_D$ (see figure 1):
\bb
\cos \tilde\theta = {\tilde{\vec x}}'_{\tilde0}\cdot\vec u_{z\tilde0}=
\,\frac{\tilde C\tilde W_{\tilde0}}{c_{\tilde0}}\,.
\ee
From the second expression (\ref{tildeW}) we obtain,
\bb
\,\frac{\tilde C\tilde W}{c}\,\sim \,\frac{ C W}{c}\lb 1+\,\frac{N^2}{A^2+B^2}
\,\frac{N}{C}\lp1-\la \frac{ C W}{c}\ra^2\rp\gamma \rb,
\ee
and
\bb
\,\frac{\tilde C\tilde W_{\tilde0}}{c_{\tilde0}}\,\sim \,\frac{ C W_{\tilde0}}{c_{\tilde0}}\lb 1+\,\frac{N^2}{A^2+B^2}
\,\frac{N}{C}\lp1-\la \frac{ C W_0}{c_0}\ra^2 \rp\gamma\rb
\sim \,\frac{ C W_{\tilde0}}{c_{\tilde0}}\lb 1+
\,\frac{N}{C}\,\gamma \rb.
\ee
We Taylor develop the leading term $w(t)\dpp=C\,W(t)/c(t)$ around $t_0$ viz\\
$w(t_0+T_D)\sim w_0+T_D\,w'_0$: 
\bb
w&=&\lp\frac{A^2+B^2}{C^2}\,\frac{c^2}{a^2}+1\rp^{-1/2}\,  ,
\\[2mm]
w'&=& \,\frac{A^2+B^2}{C^2}\,\frac{c^2}{a^2}\,  (H-H_c)\lp\frac{A^2+B^2}{C^2}\,\frac{c^2}{a^2}+1\rp^{-3/2},\qq H\dpp=\,\frac{a'}{a}\,\qq H_c\dpp=\,\frac{c'}{c}\,,
\\[2mm]
w_0&=&\cos\theta   ,\qq\qq w_0'\ =\  \sin^2\theta \,\cos\theta \,   (H_0-H_{c0}) .  
\ee
Therefore
\bb
W_{\tilde0}\,\frac{ C}{c_{\tilde0}}\sim \cos\theta\,[1+\sin^2\theta\, (H_0-H_{c0})\,T_D] ,
\ee
 and
 \bb
 \cos \tilde\theta\sim
 \cos\theta\lb1+\sin^2\theta\, ( H_0-H_{c0})\,T_D +\,\frac{\gamma}{\cos\theta}  \rb .
 \ee
The apparent angular drift of the comoving source  $\delta \theta \dpp=\tilde\theta-\theta $ then takes the form
\bb
\delta \theta\sim-\sin\theta\,\cos\theta \,(H_0-H_{c0})\,T_D-\,\frac{\gamma}{\sin\theta}\, . 
\label{dtheta}
\ee

We need a second angle to localize the direction of the incoming photons on the celestial sphere. Following standard conventions we choose the angle $\varphi $ in the $xy$-plane between the orthogonal projection of $\vec x'$ into this plane and the positive $x$-axis. Denote this projection by $\vec x'_{\perp}$ and the unit vector in the positive $x$-direction by $\vec u_{x}$,
\bb
\vec x'_\perp\dpp= \,W
 \begin{pmatrix}
 \,\frac{A}{a^2}\, \\[0.5mm]
   \,\frac{B}{a^2}\, \\[0.5mm]
   0
   \end{pmatrix} ,
   \qq
   \vec u_{x} \dpp=
\begin{pmatrix}
\frac{1}{a}\\0\\0 
\end{pmatrix}.
\ee 
Then the angle $\varphi $ today is given by
\bb
\cos \varphi = \,\frac{\vec x'_{\perp0}\cdot \vec u_{x0}}{\sqrt{\vec x'_{\perp0}\cdot \vec x'_{\perp0}}}\, = \,\frac{A}{\sqrt{A^2+B^2}}\,,
\label{defphi}
\ee
and we retrieve the usual relations of spherical coordinates:
\bb
A&=&N\cos\varphi  \,\sin\theta  ,\\
B&=&N\sin\varphi  \,\sin\theta  ,\\
C&=&N\qqq \,\cos\theta  .
\ee
Note that if $A=B=0$ the first photon moves along the $z$-axis, $\sin\theta =0$. Then the projection of the photon's velocity onto the $xy$-plane vanishes and the angle $\varphi $ is not defined. This is 
the well-known coordinate singularity of spherical coordinates at the poles.

Our last task is to compute the same angle $\tilde\varphi  $ for the second photon at its arrival time $t_{\tilde 0}=t_0+T_D$:
\bb
\cos \tilde\varphi = \,\frac{\tilde{\vec x}'_{\perp\tilde0}\cdot \vec u_{x\tilde0}}{\sqrt{\tilde{\vec x}'_{\perp\tilde0}\cdot \tilde{\vec x}'_{\perp\tilde0}}}\, = \,\frac{\tilde A}{\sqrt{\tilde A^2+\tilde B^2}}\,\sim
\,\frac{\,\frac{A}{N}\,+\alpha }{\sqrt{\,\frac{A^2+B^2}{N^2}\,-2 \,\frac{C}{N}\,\gamma }}\, 
\label{tildephi}
.
\ee
At this point we must resist a further linearization in $\alpha $ and $\gamma $.  We will understand this necessity in the next section.

\section{Linearizing the scale factors}

We know that the anisotropy of the universe cannot be large and we will concentrate on axial Bianchi I universes close to an underlying maximally symmetric one with only one scale factor $a_F(t)$, $\cdot_F$ for Friedman. It is convenient to parameterize the small deviation $\eta(t)$ in the following way:
\bb
a=\dpp a_F\,[1-{\textstyle\frac{1}{2}}  \eta],\qq
c=\dpp  a_F\,[1+ \eta].
\ee
Other linearizations are viable in Bianchi I universes \cite{stv}. This is not true in the two other axial universes that admit  maximal symmetry in the limit of small $\eta\,$: Bianchi V and IX. There the above linearization is mandatory \cite{val}. (As we are presently working on axial Bianchi IX universes and their flat limit, axial Bianchi I, I choose this parameterization here.) 

Indicating the linearization in $\eta$ and $ V$ (neglecting cross-terms $\eta V$) also by $\sim$ we have,
\bb
\eta_0=0,\qq 
H\sim H_F-{\textstyle\frac{1}{2}} \eta',\qq
H_c\sim H_F+\eta',
\ee
and
 \bb
 W&\sim& \,\frac{a_F}{N}\lb 1-{\textstyle\frac{1}{2}}\lp1-3\cos^2\theta \rp\eta\rb,
 \\[2mm]
\ul z_0+1&\sim&\,\frac{a_{F0}}{a_{Fe}}
\lb 1+ {\textstyle\frac{1}{2}} 
\lp1-3\cos^2\theta \rp\eta_e
+\cos\varphi\, \sin\theta\,V_x+\cos\theta \,V_z \rb,
\label{redshift}
\\[2mm]
 F_c&\sim\ &\frac{1}{Na_{F0}}\lp 3\,\sin^2\theta \,  \eta_e+\cos\varphi \,\sin\theta \,V_x+\cos\theta \,V_z\rp, 
 \\[2mm]
 I&\sim&\chi_{e0} +O(\eta),\qq\qq
 J\ \sim\ \chi_{e0}+O(\eta),\label{integralsIJ}
  \\[2mm]
  \alpha   &\sim&\eps_d\,(3\,\cos\varphi \,\sin\theta \, \cos^2\theta\, \eta_e+[1-\cos^2\varphi \,\sin^2\theta ] \,V_x
  \nonumber\\[2mm]
  &&\hspace{15mm}-\cos\varphi\, \sin\theta\, \cos\theta \, V_z\,), 
  \\[2mm]
  \gamma  &\sim&
  \hspace{0mm}
  -\sin\theta \,\eps_d\lp3\,\sin\theta \,\cos\theta \, \eta_e+\cos\varphi \,\cos\theta\,V_x-\sin\theta \, V_z\rp,
 \ee
 where $\chi_{e0} $
 is the conformal time of flight of the photon = the comoving geodesic `distance' between $\vec x_e$ and $\vec x_0$ in the underlying Friedman universe,
 \bb
 \chi_{e0} \dpp= \int_{t_e}^{t_0}\,\frac{1}{a_F}\, 
 \qq  \text{and}\qq
 t_{of}\sim a_{F0}\,\chi_{e0}.
   \ee
Consider first the drift with respect to $\varphi $, equation (\ref{tildephi}), and linearize it in $\eta$ and $V$, but not in $T_D/\!\!\la a_{F0}\chi_{e0}\ra
\sim\eps_d$:
\bb
 \cos\tilde\varphi  \sim 
 \frac
 {\cos\varphi  +[3\cos\varphi \,\cos^2\!\theta\,\eta_e+ 
 (1-\cos^2\!\varphi \,\sin^2\!\theta )\!\!\la V_x/\sin\theta\ra -\cos\varphi \,\cos\theta \,V_z]\,\eps_d}
 {\sqrt{1+2\cos\theta [3\cos\theta \,\eta_e+
 \cos\varphi \,\cos\theta \la V_x/\sin\theta\ra-V_z]\,\eps_d}}\,. \label{phiTlin}
\ee
   For $\sin\theta =0$, the angle $\varphi $ is ill-defined and equation (\ref{phiTlin}) is singular:
 \bb
 \lim_{\sin\theta\, \rightarrow\,0 }\cos\tilde\varphi\
 =
 \lim_{\sin\theta\, \rightarrow\,0 }
 \sqrt{\frac{1}{2\cos\varphi }\la\,\frac{V_x}{\sin\theta }\ra  \eps_d}\,.
 \ee
 If not only $V\ll 1$, but also $|\{ V_x/\sin\theta\}| \ll 1$, we may further \cite{Jonny} linearize  $\cos\tilde\varphi$:
 \bb
 \cos\tilde\varphi  &\sim& \cos\varphi \lb1+\,\frac{\sin^2\varphi }{\cos\varphi }\,\frac{V_x}{\sin\theta }\,
 \eps_d\rb.
 \ee
 In this case the drift with respect to the angle $\varphi $ is,
\bb
\delta \cos\varphi &\dpp=&\cos\tilde\varphi-\cos\varphi \sim
\,{\sin^2\varphi }\,\frac{V_x}{\sin\theta }\,\eps_d,
\label{dcosfi}\\[2mm]
\delta \varphi
&\sim&\frac{-1}{\sin\varphi }\,\, \delta \cos\varphi
 \sim-\,{\sin\varphi }\,\frac{V_x}{\sin\theta }\,\eps_d\,.
\label{dphilin}
\ee 
In particular if $V_x$ vanishes, then $\varphi $ is constant and the drift remains in the plane defined by the $z$-axis and this constant angle $\varphi $.

We encounter the well-known singularity of polar coordinates at the poles: If $\sin\theta =0$, then our source lies on the $z$-axis, there is no $\theta $-drift and the azimuthal angle  $\varphi$ is ill-defined. Now if $\sin\theta$ is small, our perturbative parameter $V_{x}/\sin\theta$ becomes large and the linear approximation fails. That means that for sources close to the $z$-axis we cannot use the $\varphi  $-drift to constrain the parameters to be fitted. We anticipate that this truncation will not reduce the statistical power of the fit significantly.

 The definition of the angle $\theta  $  does not involve the projection into the $xy$-plane and linearizing the $\theta  $-drift  is straight-forward:
 \bb
 \cos\tilde\theta  &\sim & 
 \cos\theta\lb1-\sin^2\theta\lp3 \lp{\textstyle\frac{1}{2}} a_{F0}\chi_{e0}\, \eta'_0
 +\eta_e\rp
  +\cos\varphi \,\frac{V_x}{\sin\theta}\,
 -\,\frac{V_z }{\cos\theta }\,\rp\eps_d\rb
 \hspace{-1mm} ,
 \\[2mm]
 \hspace{-4mm} \delta \cos\theta&\sim&
   -\sin^2\theta \,\cos\theta \lp3\lp{\textstyle\frac{1}{2}} a_{F0}\chi_{e0}\, \eta'_0
 +\eta_e\rp
  +\cos\varphi \,\frac{V_x}{\sin\theta}\, 
 -\,\frac{V_z }{\cos\theta }\rp\eps_d,
 \label{dcosth}\\[2mm]
\delta \theta&\sim&\lp3\sin\theta\,\cos\theta  \lp{\textstyle\frac{1}{2}} a_{F0}\chi_{e0}\, \eta'_0+\eta_e\rp
+\cos\varphi \,\cos\theta \,V_x-\sin\theta \,V_z
\rp\eps_d\,.
 \label{dthetlin}
\ee
Note that the first term, $\eta'_0$, in the drift (\ref{dthetlin}) does not depend on the redshift of the source. For a dominating expansion in the $z$-direction today, $H_{c0}-H_0\sim{\textstyle\frac{3}{2}} \eta'_0>0$,
this first term produces a drift away from the $z$-axis.

The second term, $\eta_e $, in the drift (\ref{dthetlin}) produces the opposite effect. Indeed,
for a dominating expansion in the $z$-direction from emission until today, $\eta'$ is positive and, since we have set $\eta_0$ to zero, $\eta_e$ is negative. 

If we adopt Einstein's equations with cosmological constant and dust, then $a_F$ and $\eta$ are given by,
\bb
a_F(t) = a_{F0}\lp\frac{\cosh\lp\sqrt{3 \Lambda}\, t\rp - 1}{\cosh\lp\sqrt{3 \Lambda}\, t_0\rp  - 1}\rp^{1/3},
\hspace{4mm} 
t_0={\textstyle\frac{2}{3}}\,\frac{\text{arc}\tanh\,\sqrt{\Omega _{\Lambda 0}}}{\sqrt{\Omega _{\Lambda 0}}}\, \frac{1}{H_{F0}}\,,
\hspace{4mm} 
\Omega _{\Lambda 0}\dpp=\,\frac{ \Lambda}{3\,H_{F0}^2}\,,
\label{scale}
\ee 
  and
\bb
\eta' = \eta'_0 \,\frac{a_{F0}^3}{a_F^3}\qq  
   \text{implying}\qq 
\eta(t) =
-\,{\textstyle\frac{2}{3}} \,\frac{\eta'_0}{H_{F0}} \,
\frac{\sqrt{\Omega _{\Lambda 0}}}{1-\Omega _{\Lambda 0}} 
\lb \coth\lp{\textstyle\frac{1}{2}}  \sqrt{3 \Lambda}\, t\rp-
1/\sqrt{\Omega _{\Lambda 0}}
\rb.
\label{etat}
\ee
Note that the scale factors today $a_{F0},\,a_0$ and $c_0$ are non-essential parameters and may be chosen arbitrarily. We follow the usual convention and set them equal implying $\eta(t_0)=0$. On the other hand $\eta'_0\dpp = \eta'(t_0)$ is an essential parameter also called `Hubble stretch'.

We take $a_{F0} = 1$ Gyr, $\Omega _{\Lambda 0}=0.7$ and $H_0= 72$ km/s/Mpc implying $1/H_{F0}=13.58$ Gyr, $t_0=13.09$ Gyr and $1/\sqrt{\Lambda }=9.37$ Gyr.

Let us compute the ratio $\eta_e /{(\textstyle\frac{1}{2}} a_{F0}\chi_{e0}\, \eta'_0)$ of the second term in the linearized drift equation (\ref{dthetlin}) by the first term. For the redshift $\underline z_0 = 1$, we find $t_e=5.59$ Gyr and by numerical integration $\chi _{e0}=10.48$. Finally we obtain,
\bb
\,\frac{\eta_e}{ {\textstyle\frac{1}{2}} a_{F0}\chi_{e0}\, \eta'_0}\,=-\,4.38. \label{4.38}
\ee
In contrast to Friedman universes (axial) Bianchi I universes are not isotropic. This anisotropy manifests itself via a direction-dependent Lema\^itre-Hubble `constant' (quantified by the Hubble stretch $\eta'_0$) as well as via a direction-dependent redshift for fixed emission time (quantified by $\eta_e$), equation (\ref{redshift}). The above equation (\ref{4.38}) tells us that already for a redshift $\underline z_0 = 1$ the drift is hugely dominated by the second manifestation.

This is in contradiction 
 with the result by Quercellini et al. 
\cite{Quer09}\cite{Quer10}. They claim that  -- as the redshift ranges from zero to one -- the absolute value of the ratio is at most 7 \%. They also find the opposite sign in front of the first term. 

Our independent calculation confirms the result by Marcori et al.
\cite{Mar} restricted to the axial case and to a comoving source.

\section{Conclusion}

The second term in the drift equation (\ref{dthetlin}), which is proportional to $\eta_e$ and which we find to be dominant, depends on the redshift of the source via the inverse of the direction-dependent relation (\ref{redshift}).  It is therefore phenomenologically more 
challenging  for a fit of the Gaia catalogue of quasars, 
which come with precise redshift measurements up to $\underline z_0 = 4.7$.
Both terms might help to reduce the Hubble tension.

 We look forward to this fit.

\vspace{3mm}
\noindent
{\bf Acknowledgements:} 
I thank Andr\'e Tilquin and Galliano Valent for decades of joint research and humour.
It is also a pleasure to thank Jeremy Darling for introducing me to Gaia's marvels and for his interest in my calculation. 
I am grateful to a referee and a board member of {\it Classical \& Quantum Gravity}, their feedback allowed me to improve the paper.

\appendix
  
\section{
  Noether's theorem for geodesics}

With the four Killing vectors, $\pa _x,\,\pa _y,\, \pa _z$ and $x\pa _y-y\pa _x$, Emmy Noether's theorem yields four conserved quantities on any geodesic $x^\mu (q)$. We can compute them conveniently by defining a Hamiltonian $\hh(x,p)\dpp={\textstyle\frac{1}{2}} g^{\mu \nu }(x)\,p_\mu p_\nu$ and by using Hamilton's equations with $\dot\ \dpp= {\de}/{\de q}$,
\bb
\dot x^\mu =\ \ \frac{\pa \hh}{\pa p_\mu }\,=\,p^\mu \qq\qq \text{and}\qq\qq
\dot p_\mu=-\,\frac{\pa \hh}{\pa x^\mu }\,.\label{lin2}
\ee
We interpret the first Hamiton equation  as definition of the 4-momentum $p_\mu $; then equation (\ref{lin2}) is nothing but the geodesic equation. For the conserved quantity generated by a Killing vector $\xi =\xi ^\mu \pa_\mu $ we may take $-\xi ^\mu p_\mu$. 
The four Killing vectors, then yield the four conserved quantities,
\bb 
A=-p_x =a^2\dot x,\ B=-p_y=a^2\dot y,\ C=-p_z=c^2\dot z;\  Ay-Bx=-p_xy+p_yx=0,
 \ee
 which are the familiar 3-momentum and the $z$-component of angular momentum of the point-particle moving on the geodesic trajectory $x^\mu (q)$.
 
 \section{Deriving the drift equations from Marcori et al. }

{\it Calculation by an anonymous referee of Physical Review D received on 15-10-2024}\\
\hspace{3mm}

\noindent
Consider equation (5.13) from Marcori et al. \cite{Mar}, specialized to the $z$-axis:
\bb
 \,\frac{\delta n^3}{\delta t_0}\,=-\,\frac{2}{a_0\chi _{so}}\,(\delta ^3_j -n^3n_j)(\beta ^s_j-\beta ^o_j)n^j-  (\delta ^3_j -n^3n_j)\dot\beta ^o_jn^j.
 \ee
Ref.\,\cite{Mar} considers a general Bianchi-I spacetime, while here the author specializes to the case of axial symmetry. To establish correspondence with the latter, we have to adopt the following:
\bb
\beta _1=\beta _2=-\beta _3/2=-\eta/2.
\ee
By taking 
$n^i=(\sin\theta \cos\varphi ,\sin\theta \sin\varphi ,\cos\theta )$
 and adopting $\eta^o=0$, a straightforward algebra gives
 \bb
 \delta \theta =3\sin\theta \cos\theta \lp\frac{1}{2}\,\dot\eta^o+\,\frac{\eta^s}{a_0\chi _{so}}\rp\delta t_0,
 \ee  
  which agrees with the first term in the author's eq.\,(\ref{dthetlin}). The peculiar velocity terms [second and third terms in eq.\,(\ref{dthetlin})] can be similarly shown to derive from \cite{Mar}, as the author acknowledges  in the manuscript {\it [sic]}. Thus, eq.\,(\ref{dthetlin}) already appeared in Ref.\,\cite{Mar}.

\end{document}